\documentclass[11pt,twoside]{article}

\usepackage{asp2004}
\usepackage{epsf}
\usepackage{psfig}
\usepackage{lscape}

\begin{document}
\title{Hi-Res Spectroscopy of a Volume-Limited Hipparcos Sample within 100 parsec}
\author{P. L. Lim\altaffilmark{1}, J. A. Holtzman\altaffilmark{1}, V. V. Smith\altaffilmark{2}, \& K. Cunha\altaffilmark{2}}
\altaffiltext{1}{Astronomy Department, MSC 4500, New Mexico State University, Box 30001, Las Cruces, NM 88003, USA}
\altaffiltext{2}{NOAO Gemini Science Center, Tucson, AZ, USA}

\begin{abstract}
Accurate parallaxes from the Hipparcos catalog have enabled detailed studies of stellar properties. Subgiants are of particular interest because they lie in an area where isochrones are well separated, enabling dependable age determination. We have initiated a project to obtain hi-res spectra of Hipparcos subgiants in a volume-limited sample within 100 pc. We obtain stellar properties and abundances via fully automatic analysis. We will use our results to constrain star formation history and chemical evolution in the solar neighborhood. We present initial results for our observed sample and discuss future development for this project.
\end{abstract}

\section{Sample}

\begin{figure}[!ht]
\centering
\plotone{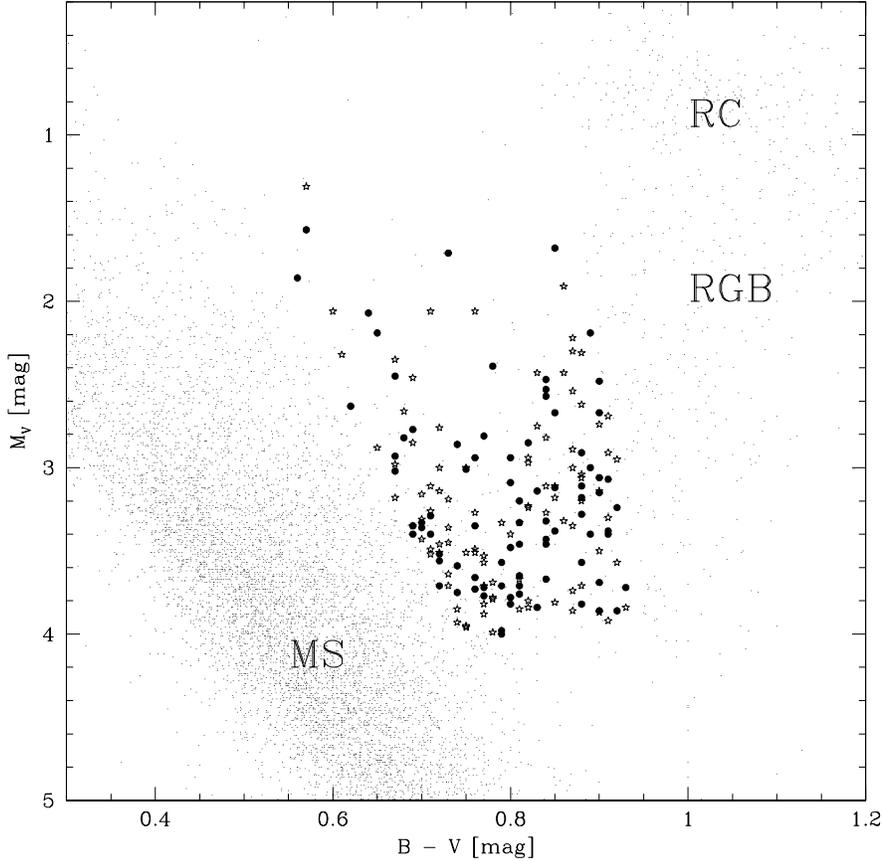}
\caption{Clean subgiants in our sample are shown as large dots (observed as of 7/18/05 UT) and stars (unobserved as of 7/18/05 UT). Small points are all Hipparcos stars within 100 pc. Other stellar populations seen here are the main sequence (MS), the red giant branch (RGB), and the red clump (RC). \label{fig-hrd}}
\end{figure}

We select Hipparcos subgiants in the Northern Hemisphere within 100 pc with $\frac{\sigma}{\pi_{o}} \leq 0.1$, as shown in Fig.~\ref{fig-hrd}. We exclude stars with known multiplicity or variability to avoid ambiguity in abundance and age analysis. A clean sample is essential to test the quality of the automated method itself. We use the ARC 3.5-m telescope at APO to obtain hi-res echelle spectra (R$\sim$37,500) for our sample. We use standard IRAF techniques for spectrum extraction and calibration. By collecting our own data, we ensure a uniform sample with high SNR and consistency in data reduction.

\section{Analysis}

We use automated gaussian fitting for clean absorption features to obtain equivalent widths. We employ MOOG \citep{sne02} and Kurucz model atmospheres \citep{kur93} in a fully automated analysis to derive effective temperature ($T_{e}$), spectroscopic surface gravity ($\log g_{s}$), microturbulent velocity ($\xi$), and iron abundance. Values of $T_{e}$, $\log g_{s}$, and $\xi$ are obtained by simultaneously solving for constant Fe {\scriptsize I} abundance across excitation potential and reduced equivalent width, and ionization balance between Fe {\scriptsize I} and Fe {\scriptsize II}. These solutions are used to obtain abundances for other elements. We use Bayesian estimation \citep{jor05} of Padova isochrones \citep{gir02} to obtain age and mass.

\section{Initial Results - Stellar Parameters \& Abundances}

We use hi-res spectral atlases of Sun and Arcturus to test the validity of our method. For Arcturus, we also have observed data to test the stability of our method at our echelle resolution. The results for these two stars agree with current standard values within errors, which implies that the method should work for our subgiants.

Disagreement between spectroscopic and trigonometric surface gravities has been reported by \citet{all99} and \citet*{tho04}, but is not well understood. In our results, we see an average scatter of $\sim$0.2 dex in the difference between spectroscopic surface gravity from our automatic method and its trigonometric counterpart from Bayesian estimation. The scatter is slightly larger at lower metallicity, which could be due to NLTE effects \citep{all99}.

\begin{figure}[!ht]
\centering
\plotone{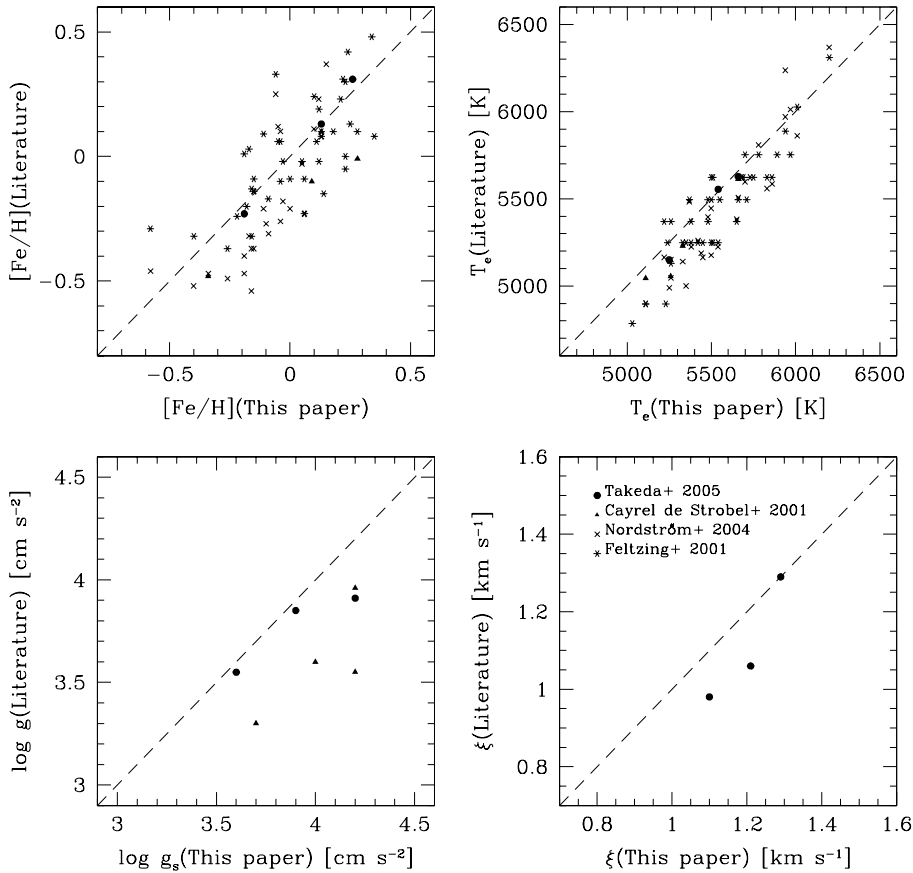}
\caption{Comparison of metallicity and stellar parameters between our method and literature. Solid symbols are values from spectroscopic studies. Crosses and asterisks are values from photometric studies. References for respective symbols are labeled in the bottom right panel. \label{fig-comp}}
\end{figure}

Comparison of our results with literature values are shown in Fig.~\ref{fig-comp}. Our values are consistent with those from \citet{tak05}, who also used fully automated analysis but with a different algorithm. It is difficult to determine the significance of our disagreement with \citet*{cay01} because their catalog is a compilation of various papers.

We have considerable scatter and apparent offsets of $\sim$0.1 dex and $\sim$100 K for [Fe/H] and $T_{e}$ respectively compared to results from photometric studies \citep*{fel01,nor04}. However these offsets are within errors. Several stars that are in both photometric studies demonstrate [Fe/H] discrepancies as large as $\sim$0.5 dex between the papers, which suggests inconsistency in photometric calibrations.

\section{Future Work}

More spectra will be acquired until we have covered all the clean subgiants in our sample. We are currently investigating the discrepancy between our data and literature values. We will improve the accuracy of our abundances by enhancing the robustness of our method. Our results will be available in the form of a catalog that is easily accessible to the scientific community.

The ultimate scientific goal of the project is to use the observed metallicity to constrain solar neighborhood SFH and chemical evolution. By incorporating observables into SFH derivation via CMD modelling, we will have results that are not only qualitatively accurate but also quantitatively.

\acknowledgements
Support for this work is provided by NSF. This research has made use of the SIMBAD database, operated at CDS, Strasbourg, France.


\end{document}